\documentclass[onecolumn,showpacs]{revtex4}

\usepackage{graphicx}
\usepackage{dcolumn}
\usepackage{bm}

\input epsf

\begin{document}

\title{\Large HOLOGRAPHIC DARK ENERGY AND VALIDITY OF THE GENERALIZED SECOND LAW OF THERMODYNAMICS
 IN THE DGP BRANE WORLD}
\smallskip

\author{\bf~Jibitesh~Dutta $^{1,3}$\footnote{jdutta29@gmail.com,~jibitesh@nehu.ac.in}
, Subenoy~Chakraborty $^2$\footnote{schakraborty@math.jdvu.ac.in}
and ~M.Ansari $^3$}

\smallskip

\affiliation{$^{1}$Department of Basic Sciences and Social
Sciences,~ North Eastern Hill University,~NEHU Campus, Shillong -
793022 ( INDIA )}

\affiliation{$^2$Department of Mathematics,~Jadavpur
University,~Kolkata-32, India.}

\affiliation{$^3$Department of Mathematics,~ North Eastern Hill
University,~NEHU Campus, Shillong - 793022 ( INDIA )}

\date{\today}

\begin{abstract}
In this paper, we investigate the validity of the generalized
second law of thermodynamics of the universe in  the DGP brane
world.The boundary of the universe is assumed to be enclosed by
the dynamical apparent horizon or the event horizon. The universe
is chosen to be homogeneous and isotropic and the validity of the
first law has been assumed here. The matter in the universe is
taken in the form of non-interacting two fluid system- one
component is the holographic dark energy model and the other
component is in the form of dust.

\end{abstract}


\maketitle


\section{\normalsize\bf{Introduction}}

 Astrophysical observations made at the turn of the last
century [1] show conclusive evidence for acceleration in the late
universe, which is still a challenge for cosmologists. It shows
beginning of accelerated expansion in the recent past.It is found
that cosmic acceleration is driven by some invisible fluid having
its gravitational effect in the very late universe. This unknown
fluid has distinguishing feature of violating strong energy
condition(SEC) being called dark energy (DE)[2].Various models
have been proposed to solve this problem. A comprehensive review
of these models is available in [3].

In the race to investigate a viable cosmological model, satisfying
observational constraints and explaining present cosmic
acceleration, brane-gravity was introduced  and brane-cosmology
was developed. A review on brane-gravity and its various
applications with special attention to cosmology  is available
in[4].

 A simple and well studied model of
brane-gravity (BG) is the Dvali-Gabadadze-Porrati(DGP) braneworld
model[5]. In this model our 4-dimensional world is a FRW brane
embedded in a 5-dimensional Minkowski bulk. It explains the origin
of DE as the gravity on the brane leaking to the bulk at large
scale.
 On the
4-dimensional brane the action of gravity is proportional to
$M_{P}^{2}$ whereas in the bulk it is proportional to the
corresponding quantity in 5-dimensions. The model is then
characterized by
 a cross over length scale
     $$ r_{c}=\frac{M_{P}^{2}}{2M_{5}^{2}}$$
such that gravity is 4-dimensional theory at scales $ a\ll r_{c}$
where matter behaves as pressure less dust but gravity
\textit{leaks out} into the bulk at scales $ a\gg r_{c} $ and
matter approaches the behaviour of a cosmological constant.

    In this conceptual set up, one of the important questions
concerns the thermodynamical behaviour of an accelerated expanding
universe driven by DE. Motivated by the profound connection
between black hole physics and thermodynamics, in recent times
there has been some deep thinking on the relation between gravity
and thermodynamics. A pioneer work in this respect was done by
Jacobson who disclosed that Einstein's
 gravitational field equation
  can be derived from the relation between horizon area and entropy together with
Clausius relation $\delta Q = T\delta S $ [6]. Some recent
discussion on the connection between gravity and thermodynamics on
various gravity theories can be found on [7]. Study of
thermodynamics in Einstein's gravity was investigated in [8].
Recently this connection between gravity and thermodynamics has
been extended to brane world scenarios [9].~In ref [10] it is
shown that apparent horizon entropy extracted through connection
between gravity and  first law of thermodynamics satisfies the
generalised second law  of thermodynamics (GSLT) in DGP warped
brane. In General Relativity (GR) frame work the authors of [11]
have shown in contrast to the case of the apparent horizon,~both
first and second law of thermodynamics breakdown if one considers
boundary of the universe to be the event horizon.But so far
attempts to address these problems are made using apparent horizon
only in brane world scenario. So it is imperative to study GSLT
using event horizon as the boundary of the universe in BG set up.

The other way to approach to the problem of DE arises from
holographic principle which states that the number of degrees of
freedom for a system within a finite region should be finite and
is bounded by the area of its boundary. As in ref [12] one obtains
holographic energy density  as
$$ \rho_{D} = 3 c^{2}M_{P}^{2}L^{-2}$$
 where $L$ is an IR cut-off in units $M_{P}^{2}=1$.
Li shows that [13] if we choose L as the radius of the event
horizon we can get the correct equation of state and get the
desired accelerating universe.

It may be noted that in literature, standard DGP model has been
generalized to (i) LDGP model by adding a cosmological constant
[14],(ii) QDGP model by adding quissence perfect fluid [15], (iii)
CDGP by Chaplygin gas [16](iv) SDGP by a scalar field [17]. In
[18] the authors analysed the DGP model by adding HDE.

    Often  using BG one obtains cosmological surprises.Thus it would
be interesting to investigate the Generalized Second Law  of
thermodynamics (GSLT) using  BG based theory.

In a recent paper[19],validity of GSLT has been studied in
Einstein's gravity and Gauss Bonnet(GB) gravity by assuming first
law of thermodynamics and conditions for validity of GSLT have
been obtained. Aim of the present paper is to extend the work of
[19] in DGP model of BG.

    Here we study the validity of GSLT  of the universe bounded by
(i) the dynamical apparent horizon (ii) event horizon in the DGP
brane world. We assume HDE density $\rho_{\Lambda} = 3
c^{2}M_{P}^{2}L^{-2}$ still holds in the DGP model and consider
the evolution of HDE on the brane according to Holographic
principle. The matter in the universe is taken in the form of
non-interacting two fluid system- one component is the holographic
dark energy  and the other component is in the form of dust(CDM).
At the apparent horizon it is shown that  GSLT is always respected
regardless of specific form of DE. But in  case bounarary is
bounded by event horizon GSLT may breakdown in the future
universe.

The paper is organized as follows : Section 2 deals with HDE in
the DGP brane model while validity of GSLT has been examined for
apparent  and event horizon  in section 3. The paper ends with a
conclusion in section 4.

\section{\normalsize\bf{HDE in the DGP model}}

In  flat, homogeneous and isotropic brane the Friedmann equation
[5] is given by
$$ H^{2}=\Big(\sqrt{\frac{\rho_{t}}{3}+\frac{1}{4r_{c}^{2}}}+\epsilon
\frac{1}{2r_{c}}\Big)^{2}       \eqno(2.1)$$
 or equivalently
$$ H^{2}-\epsilon \frac{H}{r_{c}}=\frac{\rho_{t}}{3}  \eqno(2.2)$$
 where $ H=\frac{\dot a}{a}$ is the Hubble parameter,~$\rho_{t}$ is the total cosmic fluid energy density and
  $ r_{c}=\frac{M_{P}^{2}}{2M_{5}^{2}}$ is the
crossover scale which determines the transition from 4D to 5D
behavior and $\epsilon = \pm 1 $.\\(For simplicity we are using $8
\pi G =1$ )

For $ \epsilon = 1 $, we have standard DGP(+) model which is self
accelerating model without any form of dark energy, and effective
$\omega$ is always non phantom.However for $ \epsilon = - 1 $, we
have DGP(-) model which does not self accelerate but requires dark
energy on the brane. It experiences 5D gravitational modifications
to its dynamics which effectively screen dark energy.

Here we take $\rho_{t} = \rho_{m}+ \rho_{D}$ where $\rho_{m}$ is
the energy density of CDM and $ \rho_{D} $ is the energy density
of DE.

The Friedmann eq.(2.2) can be written as
 $$ H^{2}=
\frac{1}{3}(\rho_{m}+\rho_{eff})     \eqno(2.3)$$
 where
$\rho_{eff}$ is the effective energy density given by
$$\rho_{eff}= \rho_{D} +\epsilon \frac{3H}{r_{c}}   \eqno(2.4)$$

Here we assume that there is no interaction between matter and DE.
As the two component matter system is non-interacting so they
satisfy energy conservation separately, i.e.
$$      \dot \rho_{m}+3H\rho_{m} = 0  \eqno(2.5)$$

$$ \dot \rho_{D}+3H(\rho_{D}+p_{D})=0      \eqno(2.6)$$

where $ p_{D} $ is the thermodynamic pressure of DE. Also we have
conservation equation for effective energy density[16]

 $$\dot\rho_{eff}+3H(1+\omega_{eff})\rho_{eff}=0    \eqno(2.7)$$

   where $\omega_{eff}= p_{eff}/\rho_{eff}$

   Eqs. (2.3) and (2.7) describe the equivalent GR model.
   Our choice for HDE  density is
$$\rho_{D} = \frac{3 c^{2}}{R_{E}^{2}}      \eqno(2.9) $$
where $c$ is a constant and $ R_{E}$ is the future event horizon,
given by

$$ R_{E}= a \int_{t}^{\infty} \frac{dt}{a} = a \int_{a}^{\infty} \frac{da}{Ha^{2}}
            \eqno(2.10)$$

The Friedmann eq.(2.2) can be rewritten as
$$\frac{H}{H_{0}}=\sqrt{\Omega_{m}+\Omega_{D}+\Omega_{r_{c}}}+\epsilon\sqrt{\Omega_{r_{c}}}
\eqno(2.11) $$ where

$$\Omega_{m}=\frac{\rho_{m}}{3H_{0}^{2}} ,\hspace{1.5cm}
\Omega_{D}=\frac{\rho_{D}}{3H_{0}^{2}}  , \hspace{1.5cm}
\Omega_{r_{c}}=\frac{1}{4r_{c}^{2}H_{0}^{2}}\eqno(2.12)  $$ are
the usual dimensionless density parameters and $ H_{0}$ is Hubble
parameter at redshift $z=0$.

Substituting the value of $\rho_{D}$  from (2.9) in eq. (2.10)
[18], and then differentiating the resulting equation w.r.t red
shift $z=1/a -1$, we get the evolution of $\Omega_{D}$ as
$$\frac{d \Omega_{D} }{d z}=\frac{2
\Omega_{D}^{3/2}}{c(1+z)}\Big(\frac{c}{\sqrt{\Omega_{D}}}-
\frac{1}{\sqrt{\Omega_{m}+\Omega_{D}+\Omega_{r_{c}}}+\epsilon\sqrt{\Omega_{r_{c}}}}\Big)
       \eqno(2.13)$$
       From eq(2.2), setting $z=0$ we get the initial condition of
       the above differential equation as
       $$\Omega_{D}(0)=1-2\epsilon\sqrt{\Omega_{r_{c}}}-\Omega_{m}(0)\eqno(2.13a)       $$

       We shall solve the above
       differential equation later on numerically to analyse GSLT.

       \subsection{\normalsize\bf{ Equation of State (EOS) of HDE}}
       From conservation equation (2.6) of HDE,  we get
       $$ \omega_{D}=-1+(1+z)\frac{1}{3\omega_{D}}\frac{d \Omega_{D} }{d z}    \eqno(2.14)$$
       Eliminating $\frac{d \Omega_{D} }{d z}$ from eqs. (2.13)
       and (2.14) we get
       $$ \omega_{D}=-\frac{1}{3}-\frac{2}{3c}\frac{\sqrt{\Omega_{D}}}{\sqrt{\Omega_{m}+\Omega_{D}+\Omega_{r_{c}}}+
       \epsilon\sqrt{\Omega_{r_{c}}}}    \eqno(2.15) $$
       As in GR based theory here also $\omega_{D}<-1/3$
       \begin{figure}
\includegraphics[height=2.2in]{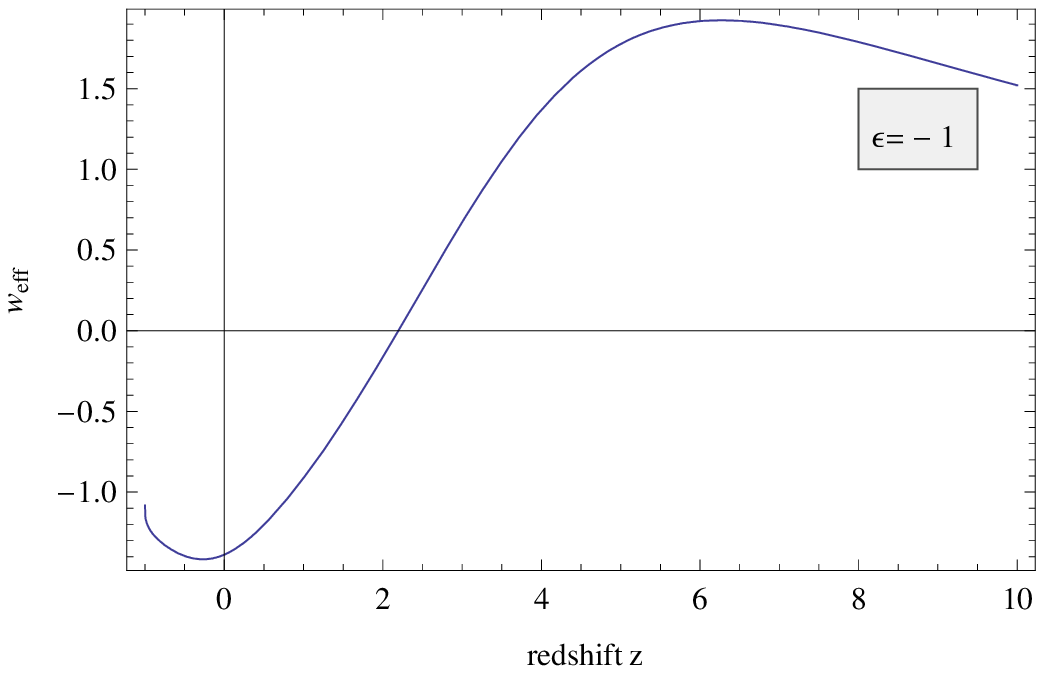}~~~~
\includegraphics[height=2.2in]{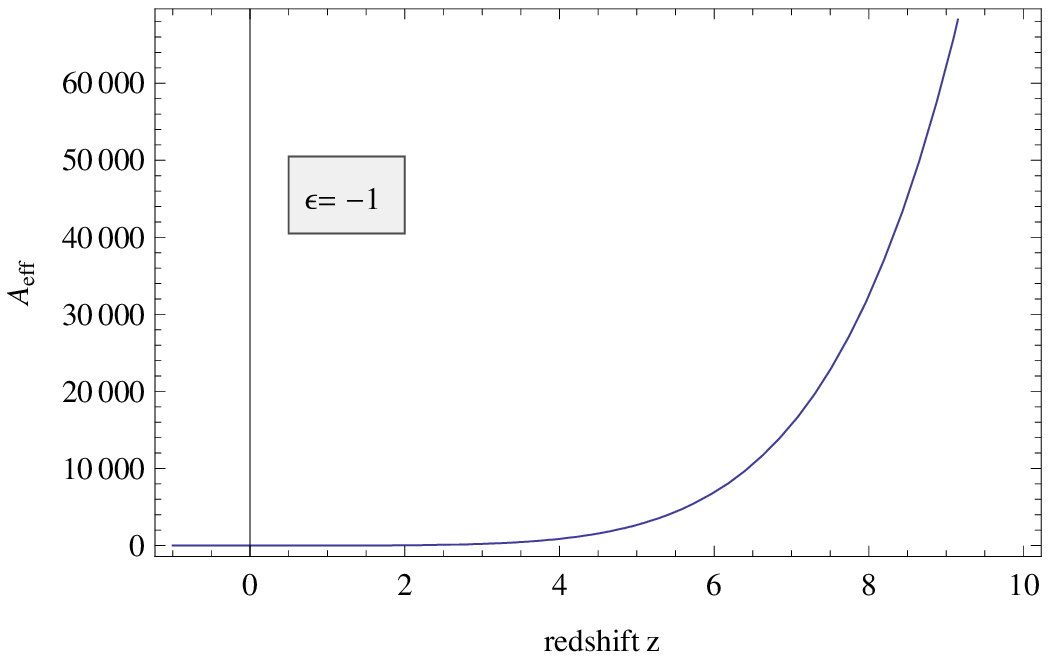}\\
\vspace{1mm} ~~~~~~~~~~~~Fig.1~~~~~~~~~~~~~~~~~~~~~~~~~~~~~~~~~~~~~~~~~~~~~~~~~~~~~~~~~~~~~~~~~~~~~~~~~~~~~~~~~~Fig.2\\

\vspace{7mm}

\includegraphics[height=2.2in]{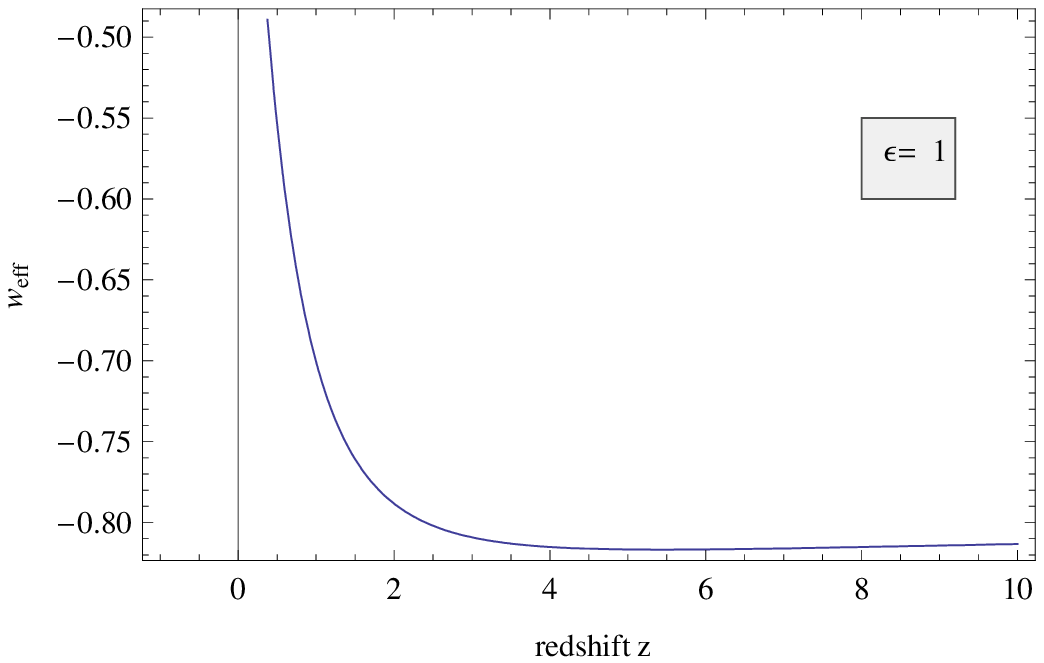}~~~~
\includegraphics[height=2.2in]{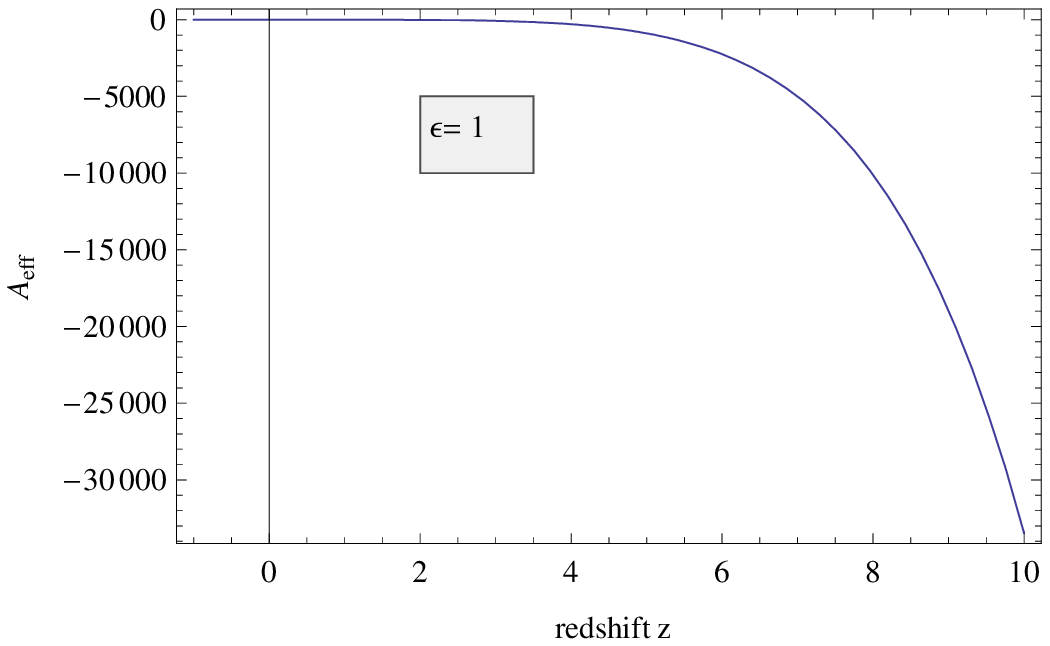}\\
\vspace{1mm} ~~~~~~~~~~~~~Fig.3~~~~~~~~~~~~~~~~~~~~~~~~~~~~~~~~~~~~~~~~~~~~~~~~~~~~~~~~~~~~~~~~~~~~~~~~~~~~~~~~~~~Fig.4\\

\vspace{7mm} {\bf Figs. 1 - 4} show the evolution of the
$\omega_{eff}$ and $A_{eff}$ w.r.t $z$. The current density
parameters used in the plots are
$\Omega_{m0}=0.3$,$~\Omega_{r_{c}}= 0.12$.

\vspace{6mm}

\end{figure}

       \subsection{\normalsize\bf{EOS of effective dark energy}}
       Defining $ \Omega_{eff}= \frac{\rho_{eff}}{3 H_{0}^{2}}$,  from eq(2.3) we can write
       $$\Big(\frac{H}{H_{0}}\Big)^{2}=\Omega_{m}+\Omega_{eff}     \eqno(2.16)$$
       Squaring eq (2.11) we have
       $$ \Big(\frac{H}{H_{0}}\Big)^{2}=\Omega_{m}+\Omega_{D}+2\Omega_{r_{c}}+
       2\epsilon\sqrt{\Omega_{r_{c}}}
       \sqrt{\Omega_{m}+\Omega_{D}+\Omega_{r_{c}}}          \eqno(2.17)  $$
       Comparing eqs(2.16) and (2.17) we  also find that
       $$\Omega_{eff}=\Omega_{D}+2\Omega_{r_{c}}+
       2\epsilon\sqrt{\Omega_{r_{c}}}
       \sqrt{\Omega_{m}+\Omega_{D}+\Omega_{r_{c}}}          \eqno(2.18)  $$
       Now from  conservation eq (2.7),we have
       $$\Omega_{eff}=\Omega_{eff}^{(0)}exp\Big(3
       \int_{0}^{z}\frac{1+\omega_{eff}(z^{'})}{1+z^{'}}dz^{'}\Big)     \eqno(2.19) $$
       Equating eqs.(2.17) and (2.18) and then taking derivative on both
       sides w.r.t  $z$, we get
       $$ 1+\omega_{eff}=\frac{1}{3\Omega_{eff}}\Big[\epsilon\sqrt{\Omega_{r_{c}}}
       \frac{3\Omega_{m}+(1+z)\frac{d\Omega_{D}}{dz}}{\sqrt{\Omega_{m}+\Omega_{D}+\Omega_{r_{c}}}
       }+(1+z)\frac{d\Omega_{D}}{dz}\Big]   \eqno(2.20) $$
       Now the above equation can be solved numerically with the
       help of differential equation (2.13) and (2.13a).The
       behavior of $\omega_{eff}$ w.r.t $z$ is shown in Figs. 1 and 3.

\section{\normalsize\bf{The Generalized Second Law of thermodynamics:}}

In this section we examine the validity of GSL on  3-DGP brane.
Let us consider a region of FRW universe
   enveloped by the  horizon and assume that the region bounded by the horizon act as a thermal
   system with boundary defined by the  horizon and is filled with a perfect fluid of energy density
    $\rho_{t}$ and pressure $p_{t}$.

We take as in[15]
$$ p_{t}= \omega_{eff}\rho_{eff} \hspace{2cm} and \hspace{2cm} \rho_{t}=\rho_{m}+\rho_{eff}   \eqno(3.1)$$

Gravity on the brane does not obey Einstein theory, therefore
usual area formula for the black hole entropy may not hold on the
brane. So we extract the entropy of the event horizon by assuming
the first law of thermodynamics on event horizon [19].

The amount of energy crossing the  horizon in time $dt$ has the
expression
$$ -dE = 4\pi R_{h}^{3}H(\rho_{t}+p_{t})dt        \eqno(3.2)  $$
where $R_{h}$ is the radius of the  horizon.

 So assuming the first law of thermodynamics we have
$$\dot S_{h} = \frac{4\pi R_{h}^{3}H}{T_{h}}\Big[\rho_{m}+(1+\omega_{eff})\rho_{eff}\Big]  \eqno(3.3)   $$
where $S_{h}$ and $ T_{h}$ are the entropy and temperature of the
 horizon respectively. Using Gibb's equation [8],
 $$ T_{h}dS_{I} = dE_{I} + p_{t}dV $$ we
obtain the variation of the entropy of the fluid  inside  the
horizon as

$$\dot S_{I} = \frac{1}{T_{h}}\Big[V\dot\rho_{t}
+(\rho_{t}+p_{t})\dot V\Big]   \eqno(3.4)$$ where $S_{I}$ and
$E_{I}$ are the entropy and energy of the matter distribution
inside the  horizon. Here we assume as in ref [20] the temperature
of the source inside the  horizon is in equilibrium with the
temperature associated with the horizon.

Connecting eqs. (2.5) and (2.7), we get
$$ \dot\rho_{t} = -3H[\rho_{m} + (1+\omega_{eff})\rho_{eff}]  \eqno(3.5)$$
So starting with $ E_{I}=\frac{4}{3}\pi R_{h}^{3}\rho_{t}$ and $ V
= \frac{4}{3} \pi R_{h}^{3}$ and using eqs.(3.4) and (3.5) and
after some simplification one gets
$$\dot S_{I} =  \frac{4\pi R_{h}^{2}}{T_{h}}\Big[\rho_{m} + (1+\omega_{eff})\rho_{eff}\Big]\Big[\dot R_{h}- HR_{h}\Big]
   \eqno(3.6) $$

Adding eqs. (3.3) and (3.6), one gets the resulting change of
entropy

$$ \dot S_{tot}=\dot S_{h}+\dot S_{I} = \frac{4\pi R_{h}^{2}}{T_{h}}\Big[\rho_{m} + (1+\omega_{eff})\rho_{eff}\Big]\dot R_{h}
 \eqno(3.7)$$

\subsection{\normalsize\bf{The dynamical apparent horizon}}
Here we study the validity of GSLT on apparent horizon.The
apparent horizon for flat space is defined as
$$ R_{A}=\frac{1}{H}$$
In terms of apparent horizon Friedmann eq.(2.3) can be written as
$$ \frac{1}{R_{A}^{2}}= \frac{1}{3}(\rho_{m}+\rho_{eff}) \eqno(3.8) $$
If we take the derivative of eq.(3.8) w.r.t cosmic time, then we
get
$$ \dot R_{A} = \frac{H R_{A}^{3}}{2}\left[\rho_{m} + (1+\omega_{eff})\rho_{eff}\right]  \eqno(3.9)$$

Substituting the above value of $\dot R_{A}$ in eq.(3.7) we get
$$ \dot S_{tot}=\dot S_{A}+\dot S_{I} = \frac{2\pi R_{A}^{5}H}{T_{A}}\left[\rho_{m} +
(1+\omega_{eff})\rho_{eff}\right]^{2} \geq 0     \eqno(3.10)  $$
It may be noted that the result is true irrespective of specific
form of DE.Thus it supports earlier investigations in GR based
different DE models like the generalized chaplygin gas [21],the
HDE  [8,22] etc.

\subsection{\normalsize\bf{The cosmological event horizon}}
It is well known that in a spatial flat de Sitter universe the
cosmological event horizon given by eq.(2.10) and the apparent
horizon coincide and $ \dot R_{E}=0$ [8].Therefore for de Sitter
space from eq(3.7) we see that $\dot S_{tot}=0$,which correspond
to reversible adiabatic expansion.

Differentiating Friedmann eq.(2.3) w.r.t cosmic time , we get
$$ \dot H=-\frac{1}{2}\left[\rho_{m} +
(1+\omega_{eff})\rho_{eff}\right] \eqno(3.11)  $$ Also from
eq.(2.4),we have
$$ \dot\rho_{eff}= \dot\rho_{D} + \epsilon\frac{3\dot H}{r_{c}}
\eqno(3.12)$$
 Connecting this equation with conservation equation
of effective energy density and eq (2.9), we get
$$ \dot R_{E}=  \frac
{R_{E}^{3}}{2c^{2}}\Big[H(1+\omega_{eff})\rho_{eff}-\frac{\epsilon}{2r_{c}}(\rho_{m}
+ (1+\omega_{eff})\rho_{eff})\Big]      \eqno(3.13)$$

Connecting eqs (3.7) and (3.13),we get
$$\dot S_{tot}=\frac{2 \pi R_{E}^{5}}{c^{2}T_{E}}\Big[H(1+\omega_{eff})\rho_{eff}-\frac{\epsilon}{2r_{c}}(\rho_{m}
+ (1+\omega_{eff})\rho_{eff})\Big]\Big[\rho_{m} +
(1+\omega_{eff})\rho_{eff}\Big]      \eqno(3.14)$$ In terms of
density parameters this equation can be rewritten as
$$\dot S_{tot}=\frac{18 \pi H_{0}^{5}
R_{E}^{5}}{c^{2}T_{E}}A_{eff}(z)    \eqno(3.15)$$  where
   $ A_{eff}(z)$  is defined as
$$A_{eff}(z)= \Big[\Omega_{m}+(1+\omega_{eff})\Omega_{eff}\Big]\Big[(\sqrt{\Omega_{m}+\Omega_{D}+
\Omega_{r_{c}}}+\epsilon\sqrt{\Omega_{r_{c}}})(1+\omega_{eff})\Omega_{eff}-
\epsilon\sqrt{\Omega_{r_{c}}}(\Omega_{m}+(1+\omega_{eff})\Omega_{eff})\Big]
\eqno(3.16) $$ Now  $  \dot S_{tot} \geq 0$ if $A_{eff}(z) \geq
0$. After solving eqs (2.13,2.13a) together with eqs
(2.18) and (2.20),the evolution $A_{eff}(z)$ is plotted in figs 2 and 4.\\
The above results lead to the following conclusions :\\

{\bf I.}  \emph{For $ \epsilon= -1$} ,~the GSLT will be valid on
the event horizon if $ (1+\omega_{eff}) > 0$ ~ i.e. effective DE
is not of the phantom nature. Then universe as a thermodynamical
system with two non-interacting fluid components(as in the present
case) always obey the second law of thermodynamics. \\

Also for late time universe or DE dominated universe
$\Omega_{m}\rightarrow 0 $ and in that case using eq.(2.11) we
have

$$A_{eff}(z)= (1+\omega_{eff})^{2}\Omega_{eff}^{2}\Big[   \frac{H}{H_{0}} +\sqrt{\Omega_{r_{c}}} \Big] \geq 0$$
Thus for DE dominated universe GSLT is always valid in this
case.\\
Further one may note that, in this case the entropy of the event
horizon also increases with time while variation of the matter
entropy with time is not positive definite, but the sum of the
entropies increases with the evolution of the universe. Also the
radius of the event horizon increases with time.\\

{\bf II.}\emph{For $ \epsilon= 1$}~ we see that $A_{eff}(z)\geq 0$
if

$$ \sqrt{\Omega_{r_{c}}} \leq \Big[\frac{\sqrt{\Omega_{m}+\Omega_{D}+
\Omega_{r_{c}}}+\sqrt{\Omega_{r_{c}}}}{\Omega_{m}+(1+\omega_{eff})\Omega_{eff}}\Big](1+\omega_{eff})\Omega_{eff}$$
Thus in this case GSLT is valid  subject to the above
inequality.\\
For late time universe or DE dominated universe
$\Omega_{m}\rightarrow 0 $ and  using eq.(2.11) we have

$$A_{eff}(z)= (1+\omega_{eff})^{2}\Omega_{eff}^{2}\Big[   \frac{H}{H_{0}} - \sqrt{\Omega_{r_{c}}} \Big]  $$
Thus for DE dominated universe GSLT is  valid in this case if  $ H
\geq  H_{0} \sqrt{\Omega_{r_{c}}}$

\section{\normalsize\bf{Conclusions:}}
In this work, we have examined the validity of generalised  second
law of thermodynamics for universe as a thermodynamical system
bounded by apparent horizon or event horizon in the DGP brane
world scenario. Assuming the validity of the first law of
thermodynamics,the GSLT is always is always satisfied on the
apparent horizon irrespective of the equation of state for non
interacting holographic dark energy and choice of $\epsilon (=\pm
1)$. For validity of GSLT on the event horizon, the choice of
$\epsilon $ as well as the equation of state for holographic dark
energy is crucial. Figures 1 and 3 show the variation of
$\omega_{eff}$  as a function of redshift $z$ for $\epsilon = -1$
and $+1$ respectively. For $\epsilon = -1$, the effective fluid
may have phantom nature around $z=0$ while for $\epsilon = +1$ the
effective fluid can not have phantom behaviour throughout the
evolution of the universe. Figures 2 and 4 show the validity of
GSLT for $\epsilon = -1$ and $+1$ respectively. For $\epsilon =
-1$,although $A_{eff}$ is a decreasing function but it is positive
throughout the evolution of the universe and hence GSLT is
satisfied on the event horizon.However,for $\epsilon =
+1$,although $A_{eff}$ is an increasing function (decreasing
function of $z$) but it is negative throughout and hence there is
a violation of GSLT.

 Moreover, from observational data the estimate of the parameters
 are the following:
 $\Omega_{m0}=0.3$,$~\Omega_{r_{c}}= 0.12$.
Taking $c=1.2$ , for $ \epsilon  =-1$,using eqs.(2.13),(2.13a) and
(2.20),we get $w_{eff}=-1.38803$ at $z=0$ and for $\epsilon  =1$
the value of $ w_{eff}=-0.157402$ at $ z=0$. Thus for $ \epsilon
=-1$, we have effective phantom behaviour
and for $ \epsilon  =1$,we have effective quintessence behaviour in the present universe. \\\\
{\bf Acknowledgement:}\\
The paper is done during a visit to IUCAA, Pune, India.The authors
are thankful to IUCAA for warm hospitality and facility of doing
research works.\\\\

{\bf References:}\\
\\
$[1]$ Perlmutter, S. J.  $et$ $al.$(1999), Astrophys. J. {\bf
517},565; (1998) astro-ph/9812133;\\D. N. Spergel $et$ $al$,
Astrophys J. Suppl. {\bf 148} (2003)175[ astro-ph/0302209] and references therein.\\\\
$[2]$Riess, A. G. $et$ $al$,(2004), Astrophys. J. {\bf 607},  665
[ astro-ph/0402512].\\\\
$[3]$Copeland,~E.J.,Sami,~M. and  Tsujikawa,~S.  Int.
J.Mod.Phys.D,
{\bf 15},(2006)1753 [hep-th/0603057] and references therein. \\\\
$[4]$Rubakov, V. A.,(2001), Phys. Usp. {\bf 44}, 871
[hep-ph/0104152];\\~Maartens, R.,(2004) Lving Rev. Relativity,
{\bf 7}, 7,  [(2003) gr-qc/0312059];\\ ~Brax, P. et al,(2004),
Rep. Prog.Phys. {\bf 67}, 2183 [hep-th/0404011]; \\~Csa$^{\prime}$ki, C.,(2004) [hep-ph/0404096].\\\\
$[5]$ Dvali,G.R. , Gabadadze,G. and Porrati, M. (2000), Phys.Lett.
{\bf B 485}  208 [hep-th/000506];\\~ Deffayet,D. (2001),
Phys.Lett. {\bf B 502}  199;\\~ Deffayet,D., Dvali,G.R. and
Gabadadze,G.(2002), Phys.Rev.{\bf D 65}  044023 [astro-ph/0105068].\\\\
$[6]$ Jacobson, T. (1995), Phys.Rev.Lett. {\bf 75} 1260
[gr-qc/9504004].
\\\\
$[7]$  Eling, C., Guedens,R. and Jacobson, T. (2006),
Phys.Rev.Lett. {\bf 96} 121301;\\~ Akbar,M. and Cai, R.G. (2006),
Phys.Lett. {\bf B 635}  .\\\\
$[8]$ Izquierdo, G. and  Pavon, D.   ,(2006), Phys. Lett. {\bf B
633},  420;\\~ Wang, B., Gong, Y. and  Abdalla, E.  (2006),
Phys.Rev.{\bf D 74}  083520
 [gr-qc/0511051];\\~Akbar,M. and Cai, R.G.   (2006), Phys.Rev.{\bf D 73}  063525
[gr-qc/0512140];\\~ Sadjadi, M. H.   (2007), Phys.Rev.{\bf D 75}
084003 [hep-th/0609128].\\\\
$[9]$ Cai, R.G. and Cao, L. M.(2007) , Nucl. Phys. {\bf B 785},
  135;\\~ Sheykhi, A. , Wang, B. and Cai, R.G. (2007) , Nucl. Phys. {\bf B 779},
  1;\\~ Sheykhi, A. , Wang, B. and Cai, R.G.  (2007), Phys.Rev.{\bf D 76}  023515 [hep-th
  /0701261];\\ ~Sheykhi, A. (2009)
  JCAP {\bf 0905}, 019 (2009).  \\\\
$[10]$ Sheykhi, A. , Wang, B. [arXiv:0811.4478]\\\\
$[11]$ ~ Wang, B., Gong, Y. and  Abdalla, E.  (2006),
Phys.Rev.{\bf D 74}  083520 [gr-qc/0511051].  \\\\
$[12]$ Cohen, A.G., Kaplan,D.B.,and Nelson,A.E. (1999),
Phys.Rev.Lett.{\bf 82} 4971.\\\\
$[13]$ Li, M. (2004), Phys. Lett. {\bf B  603},  01.\\\\
$[14]$  ~Lue,A. and ~Starkman,G.~D.(2004)
  Phys.\ Rev.\  D {\bf 70}, 101501
  [arXiv:astro-ph/0408246].   \\\\
$[15]$ ~Chimento,L.~P., ~Lazkoz, R., ~Maartens, R. and ~Quiros,
I.(2006)
  JCAP {\bf 0609}, 004  [arXiv:astro-ph/0605450].\\\\
$[16]$ ~Bouhmadi-Lopez, M. and ~Lazkoz, R.(2007)
  Phys.\ Lett.\  B {\bf 654}, 51
  [arXiv:0706.3896 (astro-ph)].\\\\
$[17]$~Zhang,~H. and Zhu,~Z.~H.(2007) Phys.Rev.{\bf D
75},023510.\\\\
$[18]$ Wu,~X, Cai,~R.~G. and Zhu,~Z.~H (2008)Phys.Rev.{\bf D
77},043502.\\\\
$[19]$ Mazumder,~N. and Chakraborty,~S.(2009) Gen Relativ
Gravit   $doi       = "10.1007/s10714-009-0881-z"$
\\\\
$[20]$Saridakis,F.N.,Gonzalez-Dyaz,P.F.and Siguenza,
C.I.[arXiv:0901.1213/astro-ph];\\~Saridakis,F.N.,Gonzalez-Dyaz,P.F.and
Siguenza, C.I. Nucl. Phys.(2004) {\bf B 697}, 363;\\~Pereira,
S.H.and Lima,J.A.S. (2008), Phys. Lett. {\bf B  669}, 266.\\\\
$[21]$ Gong,~Y.,Wang,~B. and Wang,~A. (2007) Phys.Rev.{\bf D
75},1235160. \\\\
$[22]$ Sheykhi,~A. [arXiv:0909.2427].

\end{document}